\documentclass[preprint,5p,twocolumn,number]{elsarticle}
\usepackage{amsmath,amssymb,amsfonts}
\usepackage{mathtools}
\usepackage{algorithmic}
\usepackage{subcaption}
\usepackage{graphicx}
\usepackage{textcomp}
\usepackage{xcolor}
\usepackage{todonotes}
\usepackage[]{listofsymbols}
\usepackage{siunitx}
\usepackage{tikz}
\usepackage[utf8]{inputenc}
\usepackage{pgfplots} 
\usepackage{pgfgantt}
\usepackage{pdflscape}
\pgfplotsset{compat=newest} 
\pgfplotsset{plot coordinates/math parser=false}
\usepackage{bm}
\usetikzlibrary{arrows, positioning, calc, arrows.meta, shapes.multipart, patterns, bending,}
\usetikzlibrary{plotmarks}
\usepgfplotslibrary{patchplots}
\usepgfplotslibrary{fillbetween}
\usepackage{grffile}
\usepackage{tabularx,booktabs}
\newcommand{\mb}[1]{\boldsymbol{#1}}
\usepackage{nomencl}
\usepackage{multicol}
\usepackage{framed}
\makenomenclature

\usepackage{etoolbox}
\renewcommand\nomgroup[1]{%
	\item[\bfseries
	\ifstrequal{#1}{D}{Latent Force Models}{%
	\ifstrequal{#1}{G}{Gaussian Process}{%
	\ifstrequal{#1}{T}{Thermal energy modeling}{}}}%
]}
\renewcommand*\nompreamble{\begin{multicols}{2}}
\renewcommand*\nompostamble{\end{multicols}}

\usetikzlibrary{arrows.meta}

\tikzset{%
	>={Triangle[scale width=0.75,scale=1]}, 
	shorten >=-0.25pt,	
	shorten <=-0.1pt,
	highlight/.style={rectangle,rounded corners,draw=orange,thin,inner sep=0pt,outer  sep=0pt}, 
	line join = miter, 
	abdist/.style={above=#1, anchor=base}, 
	abdist/.default=1.25mm,
	->-/.style={decoration={
		markings,
		mark=at position #1 with {\arrow{>}}},postaction={decorate}},
	->-/.default={.5},
	-<-/.style={decoration={
			markings,
			mark=at position #1 with {\arrow{<}}},postaction={decorate}},
	-<-/.default={.5},
}

\tikzset{
	frame/.style={ 
		rectangle,
		minimum size=6mm,
		thick,
		draw 
	},
	sum/.style={ 
		circle,
		minimum size =2mm,
		thick,
		draw=black, 
		inner sep = 0pt
	},
	knot/.style={ 
		circle,
		minimum size =1mm,
		fill=black, 
		inner sep=0pt, 
		line width=0pt,
	}
}

\journal{Energy and Buildings}
   
\begin{document}
	
	\begin{frontmatter}
		\title{Occupancy Prediction for Building Energy Systems with Latent Force Models%
		}
		
		\author[fau]{Thore Wietzke\corref{cor1}%
		}
		\ead{thore.wietzke@fau.de}
		
		\author[bosch]{Jan Gall}
		\ead{jan.gall@de.bosch.com}
		
		\author[fau]{Knut Graichen}
		\ead{knut.graichen@fau.de}
		
		\cortext[cor1]{Corresponding author}
		\affiliation[fau]{organization={Chair of Automatic Control, Friedrich-Alexander Universität Erlangen–Nürnberg},
			addressline={Cauerstraße 7},
			postcode={91058},
			city={Erlangen},
			country={Germany}}
		\affiliation[bosch]{organization={Corporate Sector Research and Advance Engineering,
				Advanced Energy Systems (CR/AES), Robert~Bosch~GmbH},
			addressline={Robert-Bosch-Campus~1},
			postcode={71272},
			city={Renningen},
			country={Germany}}
		
		\begin{abstract}
			This paper presents a new approach to predict the occupancy for building energy systems (BES). A Gaussian Process (GP) is used to model the occupancy and is represented as a state space model that is equivalent to the full GP if Kalman filtering and smoothing is used.
The combination of GPs and mechanistic models is called Latent Force Model (LFM). An LFM-based model predictive control (MPC) concept for BES is presented that benefits from the extrapolation capability of mechanistic models and the learning ability of GPs to predict the occupancy within the building. 
Simulations with EnergyPlus and a comparison with real-world data from the Bosch Research Campus in Renningen show that a reduced energy demand and thermal discomfort can be obtained with the LFM-based MPC scheme by accounting for the predicted stochastic occupancy. 

		\end{abstract}
	
	\end{frontmatter}
	
	\nomenclature[D]{$\bm{x}$}{state variable}
\nomenclature[D]{$\bm{z}$}{latent state variable}
\nomenclature[D]{$\bm{x_a}$}{augmented state variable}
\nomenclature[D]{$\bm{u}$}{control input}
\nomenclature[D]{$d$}{disturbance}
\nomenclature[D]{$\hat{d}$}{estimated disturbance}
\nomenclature[D]{$w$}{white noise}
\nomenclature[D]{$n_x$}{number of states}
\nomenclature[D]{$n_u$}{number of controls}
\nomenclature[D]{$n_y$}{number of outputs}
\nomenclature[D]{$n_z$}{number of latent states}

\nomenclature[D]{$n$}{number of datapoints}
\nomenclature[D]{$\bm{F}$}{latent state matrix}
\nomenclature[D]{$\bm{H}$}{latent output matrix}
\nomenclature[D]{$\bm{L}$}{noise input matrix}

\nomenclature[D]{$S(\omega)$}{power spectrum density}

\nomenclature[T]{$T_x$}{temperature of node $x$}
\nomenclature[T]{$C_x$}{thermal capacity of node $x$}
\nomenclature[T]{$R_{x,y}$}{thermal resistance between node $x$ and $y$}
\nomenclature[T]{$c$}{specific heat capacity}
\nomenclature[T]{$\dot{m}$}{mass flow}
\nomenclature[T]{$Q$}{heat flow}
\nomenclature[T]{$X$}{CO\textsubscript{2} concentration}
\nomenclature[T]{$m$}{mass}
\nomenclature[T]{$g$}{CO\textsubscript{2} generation rate}
\nomenclature[T]{$N_{Occ}$}{number of occupants}

\begin{table*}[!t]
	\begin{framed}
		\printnomenclature
	\end{framed}
\end{table*}

	\section{Introduction}

Buildings account for about 30\% of the global energy consumption~\cite{IEA2022}. A large share is used for Heating, Ventilation and Air Conditioning (HVAC) to meet the thermal comfort of occupants. In building energy systems (BES) mainly rule-based controllers (RB)~\cite{Clauss2017} are used, which specify the set points for the demand and producer side. These controllers often lack dynamic information about the BES and therefore can not exploit the thermal dynamics for efficient control. Furthermore, the HVAC system is often oversized to guarantee thermal comfort of the occupants in every circumstance.
Thus, energy-efficient control strategies have a great potential to reduce the energy demand. Besides energy efficiency, the thermal comfort of the occupants must be ensured. This works against energy efficiency, as energy must be expended to in- or decrease the temperature inside the building. 

A promising and profound alternative is model predictive control (MPC). MPC solves an optimal control problem (OCP) over a time horizon and predicts future states using the system dynamics. Furthermore, constraints and disturbances can be considered. This leads to an energy-efficient control, which inherently can consider the thermal comfort of occupants.
The major problem of MPC in BES is the model~\cite{Stoffel2023}. Since every building is different a lot of work has to be done to identify the model. If the system dynamics are known, the MPC can reduce energy demand by up to 30\% compared to RB controllers~\cite{DeConinck2016}. 

Besides unknown models, disturbances have a great impact on the performance of MPC controllers. Therefore, predicting these disturbances can further enhance the performance of MPC. Disturbances in BES are e.g. solar radiation, ambient temperature and occupancy. Predictions for weather-related disturbances can be easily retrieved from local weather forecasting services which offer a high accuracy for one-day ahead predictions. 

Besides weather effects, occupancy has a major impact on the BES. Not only do people generate heat, they also exhale CO\textsubscript{2} reducing the air quality. In addition, if occupancy is detected, the thermal comfort of occupying people has to be ensured. Thus, predicting and estimating the occupancy is a crucial task for energy efficient control and therefore is the focus of this paper.

Various occupancy estimation techniques exist which can be divided into analytical and data-driven methods~\cite{Rueda2020}.
Analytical methods model the CO\textsubscript{2} dynamics as a mass balance~\cite{Jemaa2018} and estimate the disturbance introduced by humans. To this end, the CO\textsubscript{2} generation rate of humans must be known in advance. Proposed algorithms use a finite-difference approach to approximate the time derivative of the CO\textsubscript{2} concentration~\cite{Cali2015} or a steady-state assumption to set the derivative to zero~\cite{Wang1998}.
The data-driven methods mainly use artificial neural networks \cite{Yang2014}, support vector machines~\cite{Lam2009} and hidden Markov models~\cite{Ryu2016}. Their main advantage is the higher accuracy compared to analytical methods, presumably by using more environmental data like temperature and humidity. Detrimental is the need for a large sample size of training data, which may be difficult to record due to data protection. In addition, a ground truth has to be provided for these methods, which can either be determined through surveys or cameras~\cite{Zhang2022}. An overview of various estimation techniques is shown in~\cite{Rueda2020}.

The occupancy estimation is needed for supervised prediction models which can be classified into deterministic and stochastic schedule models as well as machine learning methods~\cite{Ding2022}. Markov chains are popular stochastic models for occupancy prediction~\cite{Dong2011}. They naturally model the temporal evolution of a stochastic process and are easy to compute. Disadvantageous is their complex parameter learning and applicability.
The main machine learning methods are convolutional and recurrent neural networks~\cite{Wang2018}.
Similar to the case of occupancy estimation, the accuracy of neural networks for occupancy prediction is higher compared to deterministic or stochastic models. However, neural networks have a high demand of data to train them properly.

This work proposes a novel approach for predicting occupancy with Gaussian Processes (GP). GPs are a non-parametric machine learning method~\cite{Rasmussen2005} that can approximate general input-output relations. Prior information such as periodicity or smoothness is easily incorporated with the kernel functions of the GP. Unlike neural networks, GPs need fewer data points for training and learning. Additionally, the number of hyperparameters is much lower than in neural networks or markov chains.

GPs generally have a time complexity of $\mathcal{O}(n^3)$ for $n$ data points. There exist sparse approximations which reduce the complexity to $\mathcal{O}(nm^2)$ with $m$ inducing points~\cite{Rasmussen2005}. 
For an infinite horizon, both techniques pose to be intractable as in every time step new data is generated. If the GP purely depends on time, one can represent the GP as a state space model~\cite{Hartikainen2010}. This leverages the linear time complexity of the Kalman Filter (KF), which is used for inference and prediction. 

Combining known system dynamics with a time dependent GP, which represents an unknown force or disturbance, is called latent force model (LFM)~\cite{Alvarez2009}. This combination has the extrapolation capability of a parametric system model and the flexibility of the data-driven GP.  
The state space representation easily incorporates into LFM, as one can form an augmented state space with the known system dynamics and the data driven GP.
For BES, this augmented state space was already used in~\cite{Ghosh2015} for modeling the thermal dynamics of a building.

Using this state space model inside MPC is a promising approach. In~\cite{Saerkkae2019a}, second order linear systems were considered, thus reducing the MPC to a linear quadratic regulation (LQR) problem. As a demonstration example, a damped spring model was used. The comparison between basic LQR and LFM-LQR shows a significantly better performance of the LFM-LQR. This has been further advanced to general MPC formulations called LFM-MPC~\cite{Grasshoff2019}. Moreover,~\cite{Landgraf2022} extended this approach to nonlinear system dynamics.

In this paper, the LFM-MPC is applied to BES with occupancy as the main disturbance. In Section \ref{sec:Methods}, the general model layout is revised and the used kernel functions are introduced. Afterwards in Section \ref{sec:Occ}, the estimation and prediction of occupancy with LFM based on real world measurements from Bosch is described. Finally, two test cases are considered: a building modeled in EnergyPlus and the Bosch Research Campus in Renningen, Germany. The LFM is evaluated within an MPC, which controls the demand side of a BES.
	
	\section{Methodology}
\label{sec:Methods}

This section first explains the general problem description with a latent disturbance that is modeled by means of a Gaussian process. Finally, the representation of the Gaussian process as a state space model and its combination with MPC is outlined.

\subsection{Latent Force Model}
A continuous-time nonlinear system can be described by

\begin{equation}
	\begin{aligned}
		\dot{\mb{x}}(t) &= \mb{f}(\mb{x}(t),\mb{u}(t),d(t),t) \\
	\end{aligned}	
	\label{eq:dynamics}
\end{equation}%
with the state $\mb{x}(t) \in \mathbb{R}^{n_x}$, the inputs $\mb{u}(t) \in \mathbb{R}^{n_u}$, the measurements $\mb{y}(t) \in \mathbb{R}^{n_y}$ and the stochastic disturbance $d(t) \in \mathbb{R}$. The dynamics function $\mb{f}$ is assumed to be known.

The stochastic disturbance $d(t)$ is interpreted as a latent driving force of \eqref{eq:dynamics}. A latent variable, or function in this context, is not measured directly, thus it has to be inferred through a mathematical model. The combination of a latent function and a mechanistically model is called latent force model~\cite{Alvarez2009}. 

If $d(t)$ is not pure Gaussian white noise, a correlation between two time points $t$ and $t'$ can be assumed. Therefore, the disturbance can be modeled as a Gaussian process
\begin{equation}
	d(t) \sim \mathcal{GP}(m(t),k(t,t')).
	\label{eg:GP}
\end{equation}
with the mean function $m(t)$ and the kernel function $k(t,t')$~\cite{Rasmussen2005}. A significant feature of GPs is that they are completely described by their mean and kernel functions. Signal features like smoothness or periodicity can be incorporated into the kernel. Prediction with GP regression (GPR) comes with cubical computational complexity and quadratic memory requirements in the data points. For time-series predictions this is computationally intractable, as the number of data points grows with every time step.

If the GP is purely time-dependent, it can be transformed into a state space representation. This is done by calculating the power spectral density $S(\omega)$ of the kernel function. If the density can be written as a rational function of the form
\begin{equation}
	S(\omega) = \frac{(\mathrm{constant})}{(\mathrm{polynomial}\;\mathrm{in}\;\omega^2)}\,,
	\label{eq:SpectralDensity}
\end{equation}
one can apply spectral factorization to obtain
\begin{equation}
	S(\omega) = H(j\omega)qH(-j\omega)
\end{equation}
with the transfer function $H(j\omega)$ and the spectral density $q$ of a white noise process \cite{Hartikainen2010}. This transfer function can be transformed to the canonical state space form

\begin{align}
\begin{aligned}
			\dot{\mb{z}}(t) &= \underbrace{\begin{bmatrix}
				0 & 1 & 0 & \dots & 0 \\
				0 & 0 & 1 & \dots & 0 \\
				\vdots & \vdots & \vdots & \ddots & \vdots \\
				0 & 0 & 0 & \dots & 1 \\
				a_1 & a_2 & a_3 & \dots & a_{n_z} \\
		\end{bmatrix}}_{:=\mb{F}} \mb{z}(t) + \underbrace{\begin{bmatrix}
				0 \\
				0 \\
				\vdots \\
				0 \\
				1 \\
		\end{bmatrix}}_{:=\mb{L}} w(t) \\
		\hat{d}(t) &= \underbrace{\begin{bmatrix}
				1 & 0 & \dots & 0 
		\end{bmatrix}}_{:=\mb{H}} \mb{z}(t)
\end{aligned}
\label{eq:latentSS}%
\end{align}%
with latent states $\mb{z}(t) \in \mathbb{R}^{n_z}$, Gaussian white noise $w(t)$ and the output $\hat{d}(t)$. This white noise driven process can be viewed as a continuous hidden markov chain model~\cite{Bishop2006}. 
The latent states are inferred from a Kalman filter with a computational complexity of $\mathcal{O}(n_z^3n)$ with $n_z$ latent states and $n$ data points~\cite{Hartikainen2010}. Therefore, the KF is more efficient than the GP implementation, if the state dimensionality is considerably smaller than the number of data points. This computational demand can be reduced further, if long time-series in an infinite horizon are evaluated. Then, a time-invariant Kalman Filter can be used, whose computational complexity scales with $\mathcal{O}(n_z^2n)$~\cite{Solin2018}.

The dynamical system \eqref{eq:dynamics} can be augmented by the latent state space model \eqref{eq:latentSS} to allow for a joint state estimation~\cite{Landgraf2022}. The augmented state space model is given by 
\begin{equation}
\begin{aligned}
		\dot{\mb{x}}_a(t) &= \begin{bmatrix}
		\mb{f}(\mb{x}(t),\mb{u}(t),\mb{H}\mb{z}(t),t) \\
		\mb{F}\mb{z}(t)
	\end{bmatrix} + \begin{bmatrix}
		0 \\
		\mb{L}w(t)
	\end{bmatrix} \\
	&= \mb{f}_a(\mb{x}_a(t),\mb{u}(t),t) \\
\end{aligned}
\label{eq:augState}
\end{equation}
with the augmented state $\mb{x}_a = \left[\mb{x}^T, \mb{z}^T \right]^T $. The latent state space is essentially a disturbance model derived by GPR. 

\subsection{Converting Kernel Functions} 
The latent state dimension and the corresponding state space matrices depend on the chosen kernel. A common kernel function for the GP model \eqref{eg:GP} is the Whittle-Matérn kernel 
\begin{equation}
	k_{M}(t,t') = \sigma^2 \frac{2^{1-\nu}}{\Gamma(\nu)} \left(\frac{\sqrt{2\nu}|t-t'|}{l}\right)^\nu 
	K_\nu\left(\frac{\sqrt{2\nu}|t-t'|}{l}\right) 
\end{equation}
with the parameter $\nu$ defining the smoothness and $\sigma^2$ the magnitude of the process. $\Gamma$ and $K_\nu$ are the gamma function and modified Bessel function of the second kind. For $\nu = \frac{1}{2} + p,\; p \in \mathbb{N}_0$, the spectral density of the Matérn kernel is of the form \eqref{eq:SpectralDensity}~\cite{Hartikainen2010} and therefore can be transformed into an exact state space representation with the order $p+1$. 

If the spectral density is not a rational function, no exact state space can be computed. Thus, $S(\omega)$ has to be approximated. One example is the Squared Exponential (SE) kernel~\cite{Rasmussen2005}, which can be approximated with a Taylor series or Padé approximation~\cite{Hartikainen2010}~\cite{Saerkkae2014}. The SE kernel is often the first choice for GPR, but the infinite smoothness takes a rather strong assumption on real physical processes.

Another example for non-rational spectral density is the periodic kernel
\begin{equation}
	k_{P}(t,t') = \sigma^2 \exp\left(-\frac{2\sin^2\left(\omega_0\frac{t-t'}{2}\right)}{l^2}\right) 
\end{equation}
with the magnitude $\sigma^2$, the length scale $l$ and the frequency $\omega_0$ \cite{MacKay1998}. The spectral density is non-rational because the kernel consists of an infinite number of resonators. One has to truncate the number of resonators to form a finite state space, thus the number of resonators is also the degree of approximation $J$. A detailed explanation on how to convert the periodic kernel is given in~\cite{Solin2014}. 

One important kernel for this work is the quasi-periodic kernel which allows deviations from exact periodicity. As new kernels can be constructed by multiplication and addition of existing kernels \cite{Rasmussen2005}, the quasi-periodic kernel is given by $k_{QP}(t,t') = k_P(t,t')\cdot k_{Q}(t,t')$ where $k_Q(t,t')$ can be chosen as e.g. the Matérn or SE kernel. The state transition matrix of the quasi-periodic kernel is obtained by
\begin{equation}
	\mb{F}_{QP} = \mb{F}_P \otimes \mb{I}_{n_Q} + \mb{I}_{n_P} \otimes \mb{F}_Q
	\label{eq:productQP}
\end{equation}
where $\otimes$ denotes the Kronecker product and $\mb{I}_x$ the identity matrix with dimension $x$ \cite{Solin2014}. The resulting state dimension is $n_{QP} = n_{P}\cdot n_{Q}$. Further information on converting kernel functions can be found in \cite{Saerkkae2019}. Efficient implementations for the state space representations are available in the Bayes-Newton Python package \cite{wilkinson2023bayes}.

\subsection{LFM-based Model Predictive Control}

Model predictive control is an optimization based control algorithm, which solves an optimal control problem for every time step. Using \eqref{eq:dynamics}, the OCP for the MPC with the prediction horizon $T>0$ is
\begin{subequations}
	\begin{align}
		\min_{\mb{u}}\; &J(\mb{u}) = V(\mb{x}(T)) + \int_{t_k}^{t_k+T}l(\mb{x}(\tau),\mb{u}(\tau))\,\mathrm{d}\tau \\[3pt]
		\mathrm{s.t.}\;\; &\dot{\mb{x}}(\tau) = \mb{f}(\mb{x}(\tau),\mb{u}(\tau),\hat{d}(\tau),\tau), \quad \mb{x}(t_k) = \mb{x}_{t_k}, \label{eq:OCP_f} \\
		&\mb{u}(\tau) \in \left[\mb{u}_{min},\mb{u}_{max}\right], \\
		&\mb{h}(\mb{x}(\tau),\mb{u}(\tau), \hat{d}(\tau)) \leq \mb{0}, \: \forall\tau \in \left[ t_k,t_k+T\right] \label{eq:OCP_h}
	\end{align}
	\label{eq:OCP}%
\end{subequations}%
with $t_k = t_0 + k\Delta t$ as the current time step with sampling time $\Delta t$, integral cost function $l$, terminal cost $V$ and the vector inequality constraints \eqref{eq:OCP_h}. The states for every time step are either provided through measurements or through state estimation. 

There are two ways to incorporate the disturbance $\hat{d}(t)$ into \eqref{eq:OCP_f}: using the augmented state space model \eqref{eq:augState} or pre-computing $\hat{d}(t)$. The augmented state space has the disadvantage of a higher state dimension, which results in higher computational demand. The latent state space is not controllable but observable~\cite{Saerkkae2019a}. Therefore, no change in the disturbance trajectory between optimization steps is expected. Thus, only the output $\hat{d}(t)$ of the latent state space \eqref{eq:latentSS} has to be provided over the prediction horizon. 

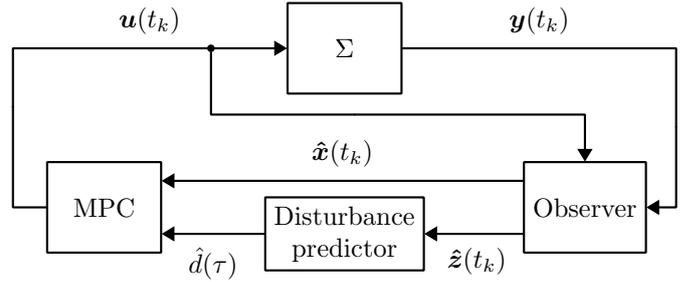
\begin{figure}
	\centering
	\begin{tikzpicture}[every path/.style={draw,thick, line  join=bevel},on grid,
		node distance=12mm and 12mm, frame/.append style={minimum width=15mm, minimum height = 6mm}]%
		
		\node(System)[frame, minimum height=12mm] at (0,0){$\Sigma$};
		\node(helper)[below= 6em of System]{};
		\node(Knot)[knot, left= 5em of System]{};
		
		\node(helperU)[below= 2.5em of System]{};
		
		\node(Dist)[frame, below= 1em of helper, align=center, yshift=-0em]{Disturbance\\predictor};
		\node(MPC)[frame, left= 9em of helper, minimum height=12mm]{MPC};
		\node(Observer)[frame, right= 9em of helper, minimum height=12mm]{Observer};
		
		\node(hMPC)[above left=of MPC]{};
		\node(hObserver)[above right=of Observer]{};
		
		\draw[-] (MPC) -| (hMPC.north);
		\draw[->] (hMPC) |- (System) node[pos=0.75, above=0.15em]{$\bm{u}(t_k)$};
		
		\draw[-] (System) -| (hObserver.south) node[pos=0.25, above=0.15em]{$\bm{y}(t_k)$};
		\draw[->] (hObserver) |- (Observer);
		
		\draw[->] ($(Observer.west) + (0,1em)$) -- ($(MPC.east) + (0,1em)$) node[pos=0.5, above=0.15em]{$\bm{\hat{x}}(t_k)$};
		\draw[->] ($(Observer.west) - (0,1em)$) -- (Dist) node[pos=0.47, below=0.15em]{$\bm{\hat{z}}(t_k)$};
		\draw[->] (Dist) -- ($(MPC.east) - (0,1em)$) node[pos=0.5, below=0.15em]{$\hat{d}(\tau)$};
		\draw[-] (Knot) |- (helperU.east);
		\draw[->] (helperU) -| (Observer);
		
	\end{tikzpicture}
	\caption{Exemplary control loop for the LFM-MPC. The observer provides state estimates for the mechanistic model and the latent states, which are used to compute the estimated disturbance trajectory $\hat{d}(\tau)$.}
	\label{fig:MPC_Loop}
\end{figure}

The resulting control scheme is shown in Figure \ref{fig:MPC_Loop}. The MPC computes the control input which is fed into the system $\Sigma$. With the system response, an observer is used to estimate the states $\bm{\hat{x}}$ and latent states $\bm{\hat{z}}$ with the augmented state space model \eqref{eq:augState}. The latent state estimate is fed into a disturbance predictor which computes the predicted trajectory for $\hat{d}(\tau)$ with $\tau \in \left[t_k, t_k+T \right] $. 

	\section{Occupancy Estimation and Prediction in BES}
\label{sec:Occ}

In this section, the demand side of a BES is presented to show the impact of occupancy for the thermal dynamics of one thermal zone. Afterwards, the occupancy estimation based on CO\textsubscript{2} measurements is outlined. Finally, a GP is used to learn the occupancy pattern. Furthermore, the resulting state space representation is compared with the full GP.

\subsection{BES Demand Side}
A full BES consists of the producer and the demand side. The producer side consists of HVAC equipment, which provides the building with the energy to meet its climate conditions. On the other hand, the demand side uses this energy in form of e.g. chilled air or heated water to meet its climate requirements. Throughout this paper, only the demand side is considered and ideal HVAC equipment is assumed.

The demand side consists of thermal zones, which describe rooms with similar comfort requirements and disturbances. We are interested in the impact of occupancy on the temperature in each zone. The dynamical model for one thermal zone is given by \cite{MassaGray2016}
\begin{subequations}
	\begin{align}
		C_z\dot{T}_z &= \frac{T_w-T_z}{R_{z,w}}\!+\!\frac{T_r-T_z}{R_{z,r}}\!+\!c_a \dot{m}_a (T_S-T_z)\!+\! Q_{occ}N_{occ} \label{eq:GB_Zone_Tz}\\
		C_w\dot{T}_w &= \frac{T_z-T_w}{R_{z,w}} + \frac{T_a-T_w}{R_{w,a}} \\
		C_r\dot{T}_r &= \frac{T_z-T_r}{R_{z,r}} +  c_w \dot{m}_w (T_V-T_r)
	\end{align}%
	\label{eq:GB_Zone}%
\end{subequations}%
with $T_z$ as the zone temperature, the external wall temperature $T_w$, the radiator temperature $T_r$, the ambient temperature $T_a$ and $T_S$ and $T_V$ for the respective supply temperature of air and water. The parameters $C_z$, $C_w$ and $C_r$ are the corresponding heat capacities for each temperature node and $R_{z,w}$, $R_{z,r}$ and $R_{w,a}$ are the heat resistances between the nodes. The specific heat capacities of air and water are $c_a$ and $c_w$. The controllable mass flow of air and water is given by $\dot{m}_a$ and $\dot{m}_w$. Note that interactions with other thermal zones can be considered in \eqref{eq:GB_Zone_Tz} by additional coupling terms. 

The heat flux through occupancy is described by $Q_{occ}N_{occ}$, where $N_{occ}$ is the number of occupants and $Q_{occ}$ the heat gain per occupant. The considered rooms are assumed to be offices, for which \cite{ASHRAE_2017} recommends a heat gain of \qty{120}{\watt} per occupant. Because of office work, this heat gain is increased by \qty{75}{\watt} per person to account for electrical equipment like monitors or laptops.

\subsection{Occupancy Estimation}
Occupancy not only impacts the temperature, as seen in the previous section, but also the CO\textsubscript{2} concentration in the considered zone. The dynamics of the CO\textsubscript{2} concentration in one thermal zone is modeled as \cite{Jemaa2018}
\begin{equation}
	\dot{X} = \dot{m}_a(X_s-X)\frac{1}{m_z} + \frac{gN_{Occ}}{m_z}
	\label{eq:CO2}
\end{equation}
with the CO\textsubscript{2} concentration $X$ in the zone, the CO\textsubscript{2} concentration $X_s$ of the supply air, the air mass of the zone $m_z$ and the CO\textsubscript{2} generation rate $g$ per human. Based on \eqref{eq:CO2}, the joint estimation of $gN_{Occ}$ and $X$ can be performed using the Unscented Kalman Filter (UKF). 

Note that the CO\textsubscript{2} generation rate is usually not constant; rather, it depends on the activity level of humans. The activity level is given either in \unit{\watt/\square\meter} or metabolic rate $\qty{1}{met} = \qty{58.1}{\watt/\square\meter}$. For typical office work, an activity level of $\qty{1.2}{met}$ can be assumed~\cite{ASHRAE_2017}. 
According to~\cite{ASHRAE} and~\cite{EplusEng}, a CO\textsubscript{2} generation rate of
\begin{equation}
	g = \qty{0.31}{\liter/\minute}\cdot \qty{1.2}{\kilogram/\cubic\meter} = \qty{6.2e-6}{\kilogram/\second}
\end{equation}
is obtained. The activity level is assumed to be constant for this paper, thus the CO\textsubscript{2} generation rate is also constant.

\begin{figure}
	\centering
%
%
\definecolor{mycolor1}{rgb}{0.00000,0.44700,0.74100}%
\begin{tikzpicture}

\begin{axis}[%
width=7.2cm,
height=4cm,
at={(0,0)},
scale only axis,
xmin=737446,
xmax=737458,
xtick={737446,737448,737450,737452,737454,737456,737458},
xticklabels={{Jan 21},{Jan 23},{Jan 25},{Jan 27},{Jan 29},{Jan 31},{Feb 02}},
xticklabel style={font=\footnotesize},
scaled x ticks=false,
ymin=0,
ymax=10,
ylabel style={font=\color{white!15!black}, font=\footnotesize},
ylabel={Occupancy},
axis background/.style={fill=white},
xmajorgrids,
ymajorgrids
]
\addplot [color=mycolor1, line width=1.5pt, forget plot]
  table[]{Occupancy.tsv};
\end{axis}

\end{tikzpicture}%
	\caption{Occupancy estimation based on measurement data from Jan-Feb 2019 from zone 21 at the Bosch Research Campus.}
	\label{fig:Occ}
\end{figure}
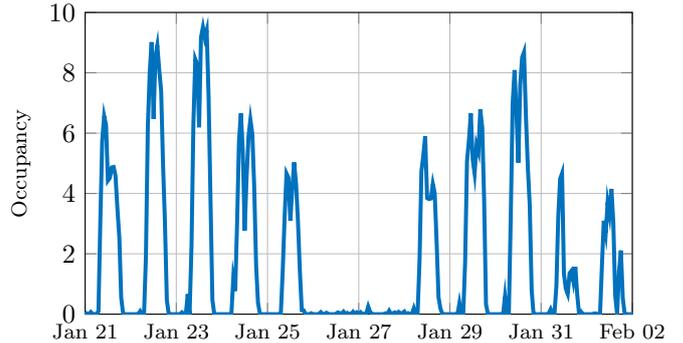

With these assumptions, the occupancy can be estimated with the UKF from measured CO\textsubscript{2} concentrations. The data was provided by Bosch for a floor of a R\&D building at their Research Campus in Renningen and was sampled with $\Delta t = \qty{15}{\minute}$. A floor plan and description is provided in Section \ref{sec:Eval}.
A sample from the estimated occupancy data is shown in Figure \ref{fig:Occ} for zone 21.

\subsection{Occupancy Prediction with GP}

Based on the UKF estimation, a GP as outlined in Section \ref{sec:Methods} is used to learn the occupancy. As one can see in Figure \ref{fig:Occ}, a daily periodic pattern is observable. This periodicity is not exact, as there are daily variations in the occupancy. The pattern is rather rough, because the occupancy can change very fast, e.g. people leaving the room. Furthermore, a similarity of the weekdays with themselves is apparent. 

This prior knowledge can be expressed in the mean and kernel function. The mean function is set to zero since no clear mean is observable. The day periodicity and the daily variations are modeled by a quasi-periodic kernel. The damping part of the quasi-periodic kernel is expressed by a Matérn kernel with $\nu=1/2$, because of the roughness of the data. In addition, the state space representation of the Matérn kernel with $\nu=1/2$ is exact and has a dimension of one. The resulting kernel is given by
\begin{equation}
	k_{QP}(t,t') = k_{periodic}(t,t')\cdot k_{Mat\acute{e}rn}(t,t').
\end{equation}
For the weekday similarity individual GPs are used, which corresponds to five individual GPs. Saturday and Sunday are not included, as no occupancy is present during the weekend.

\begin{figure}[h]
	\centering
	\definecolor{mycolor1}{rgb}{0.00000,0.44700,0.74100}%
\definecolor{mycolor2}{rgb}{0.85000,0.32500,0.09800}%
\newcommand{\MyWidth}{0.42\textwidth}
\newcommand{\MyHeight}{4cm}
\begin{tikzpicture}
\begin{axis}[%
	width=\MyWidth,
	height=\MyHeight,
	at={(0,0)},
	scale only axis,
	xmin=0,
	xlabel style={font=\color{white!15!black}, font=\footnotesize},
	ymin=-2.5,
	ymax=6.5,
	xmax = 47,
	ylabel style={font=\color{white!15!black}, font=\footnotesize, yshift=-0.75em},
	ylabel={Occupancy},
	axis background/.style={fill=white},
	title style={font=\bfseries},
	title=GP prediction,
	xmajorgrids,
	xminorgrids,
	ymajorgrids,
	legend columns=3,
	legend style={at={(0.65,-0.27)}, anchor=south, legend cell align=left, align=left, draw=white!15!black} 
	]
	\addplot [color=mycolor1, line width=1.5pt]
		table[]{GP_prediction-1.tsv};
	\addlegendentry{True}
	
	\addplot [color=mycolor2, line width=1.5pt]
		table[]{GP_prediction-2.tsv};
	\addlegendentry{Pred}
	
	\addplot [name path=A,color=black,dashed]
		table[]{GP_prediction-3.tsv};
	\addlegendentry{Mean $\pm\sigma$}
	
	\addplot [name path=B,color=black, dashed, forget plot]
		table[]{GP_prediction-4.tsv};
	\addplot [gray!40] fill between[of= A and B];
	
	\addplot[color=black, line width=1.5pt, dashdotted] 
		table[row sep = crcr]{23.92 -2.5 \\ 24 6.5 \\};
\end{axis}

\begin{axis}[%
	width=\MyWidth,
	height=\MyHeight,
	at={(0,-6cm)},
	scale only axis,
	xmin=0,
	xlabel style={font=\color{white!15!black}, font=\footnotesize},
	xlabel={t/h},
	ymin=-2.5,
	ymax=6.5,
	xmax = 47,
	ylabel style={font=\color{white!15!black}, font=\footnotesize, yshift=-0.75em},
	ylabel={Occupancy},
	axis background/.style={fill=white},
	title style={font=\bfseries},
	title=State space prediction,
	xmajorgrids,
	xminorgrids,
	ymajorgrids,
	legend columns=3,
	]
	\addplot [color=mycolor1, line width=1.5pt]
		table[]{GP_prediction-5.tsv};
	
	\addplot [color=mycolor2, line width=1.5pt]
		table[]{GP_prediction-6.tsv};
	
	\addplot [name path=A,color=black,dashed]
		table[]{GP_prediction-7.tsv};
	
	\addplot [name path=B,color=black, dashed, forget plot]
		table[]{GP_prediction-8.tsv};
	\addplot [gray!40] fill between[of= A and B];
	
	\addplot[color=black, line width=1.5pt, dashdotted] 
		table[row sep = crcr]{23.92 -2.5 \\ 24 6.5 \\};
\end{axis}

\end{tikzpicture}%
	\caption{Comparison of the GP and the state space approximation~\eqref{eq:latentSS}. The dash dotted vertical line shows the prediction start.}
	\label{fig:GP_pred}
\end{figure}
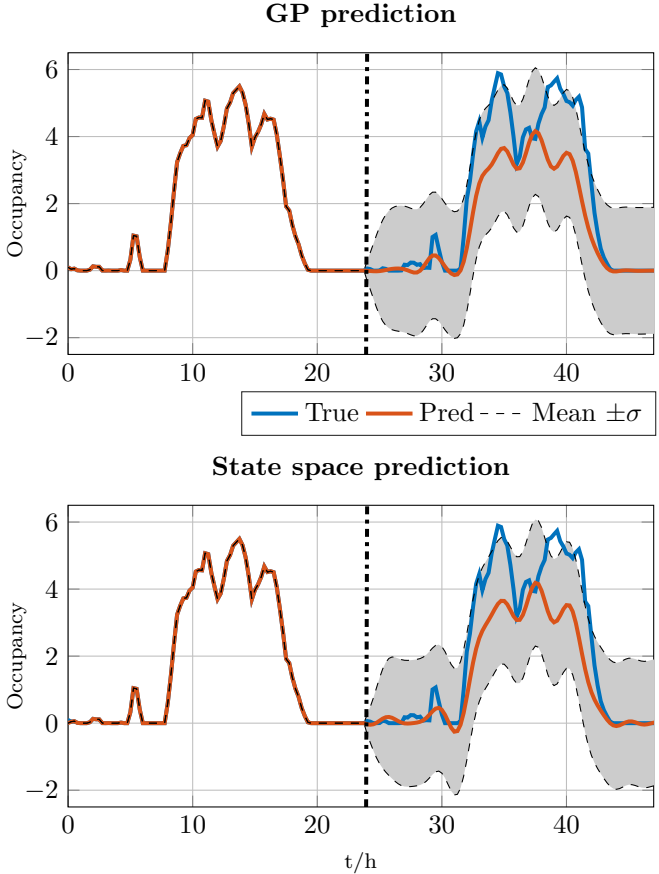

A comparison between the GP and its state space representation with experimental occupancy data is shown in Figure \ref{fig:GP_pred}. 
The data from $t=0$ to $t=\qty{24}{\hour}$ (up to the dash dotted line) is used as a training set for the GP. For the state space representation, a Kalman filter is used to initialize the states. This is equivalent to computing the covariance matrix for the GP. In the second time interval $t \in \{\qty{24}{\hour}, \qty{48}{\hour}\}$, the prediction is shown in red compared to the true occupancy in blue.

It can be seen that both exhibit nearly the same mean and variance. Note that the state space representation has to be approximated, as the power spectral density of the periodic kernel is not of the form \eqref{eq:SpectralDensity}. For this comparison an approximation degree of $J=10$ is used, which results in a state dimension of $n_{QP} = 22$.

\begin{figure}[b]
	\centering
	\definecolor{mycolor1}{rgb}{0.00000,0.44700,0.74100}%
\definecolor{mycolor2}{rgb}{0.85000,0.32500,0.09800}%
\newcommand{\MyWidth}{0.35\textwidth}
\newcommand{\MyHeight}{2cm}
\begin{tikzpicture}
\begin{axis}[%
	width=\MyWidth,
	height=\MyHeight,
	at={(0,0)},
	scale only axis,
	xmin=0,
	xlabel style={font=\color{white!15!black}, font=\footnotesize},
	xlabel={t/d},
	ymin=0.03,
	ymax=0.06,
	xmax = 3,
	ylabel style={font=\color{white!15!black}, font=\footnotesize},
	ylabel={Variance},
	axis background/.style={fill=white},
	xmajorgrids,
	xminorgrids,
	ymajorgrids
	]
	\addplot [color=mycolor1, line width=1.5pt]
		table[]{variance.csv};
\end{axis}

\end{tikzpicture}%
	\caption{Variance of the occupancy for example data for three days.}
	\label{fig:GP_var}
\end{figure}
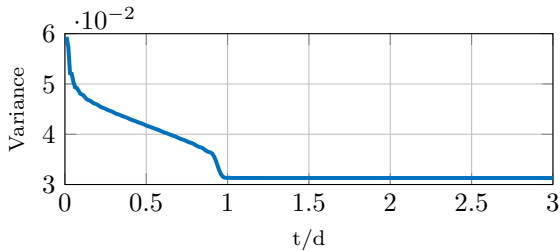

As said in Section \ref{sec:Methods}, a stationary KF can further reduce the computational demand of the occupancy prediction. As seen in Figure \ref{fig:GP_var}, the variance converges after one day, which justifies to use stationary Kalman filtering for the occupancy prediction.

	\section{Evaluation}
\label{sec:Eval}

In this section, the derived disturbance model for the occupancy is used inside an MPC to control the demand side of a BES. Two case studies are examined to show the capabilities of the proposed approach. First, an EnergyPlus model of a simple three-zone office building is considered. This leverages known occupancy data to show the best-case scenario. The second test case is based on real measurement data from the Bosch Research Campus in Renningen, Germany. 

\subsection{Test Case 1: EnergyPlus}

\begin{figure}
	\centering
	\includegraphics[width=0.45\textwidth]{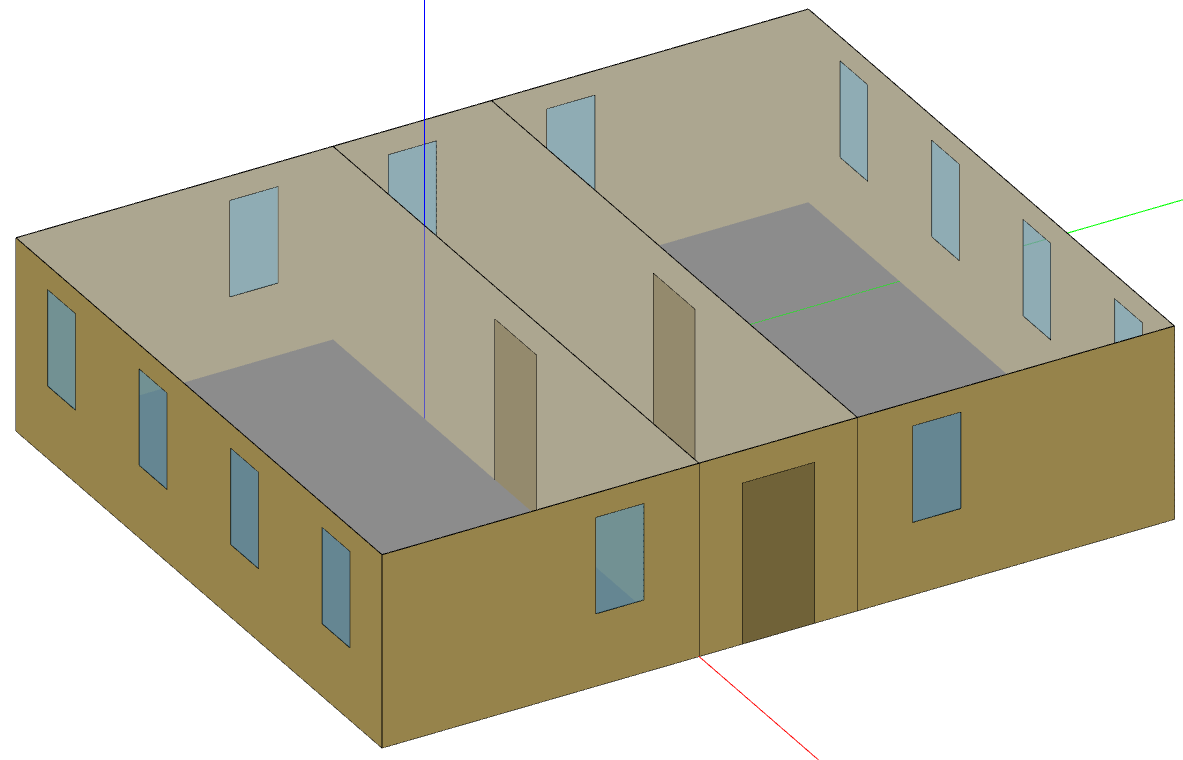}
	\caption{3D view of the building used for the EnergyPlus test case.}
	\label{fig:EnergyPlus}
\end{figure}

The EnergyPlus model consists of two offices in north and south orientation and a hallway in between. A 3D picture of the zone layout is shown in Figure \ref{fig:EnergyPlus}. Each office is equipped with a radiator and the air is supplied by an ideal HVAC system. The occupancy data was generated using normal distributions for the arrival time and duration of stay. A binomial distribution is used to model the presence of a person.

As EnergyPlus simulates the whole building with heat fluxes between each zone, zone couplings were considered. Therefore each zone was modeled by \eqref{eq:GB_Zone} with additional terms of the form
\begin{equation}
	Q_i = \frac{T_i-T_z}{R_i}
\end{equation}
in \eqref{eq:GB_Zone_Tz}. As before, $T_z$ is the corresponding zone temperature, $T_i$ the temperature of the neighbour zone and $R_i$ is the heat resistance between the two zones. The parameters were determined with the MATLAB System Identification Toolbox. The resulting model was used inside the MPC. 

EnergyPlus uses typical weather data for simulation, which include a multitude of disturbances, like solar radiation and outdoor temperature. Typical weather data from Munich was used in EnergyPlus, which are available at the website from EnergyPlus. The MPC only considered the ambient temperature. The hyperparameters for the LFM were determined manually.

\subsection{Test Case 2: Renningen}

\begin{figure*}
	\centering
	\includegraphics[width=\textwidth]{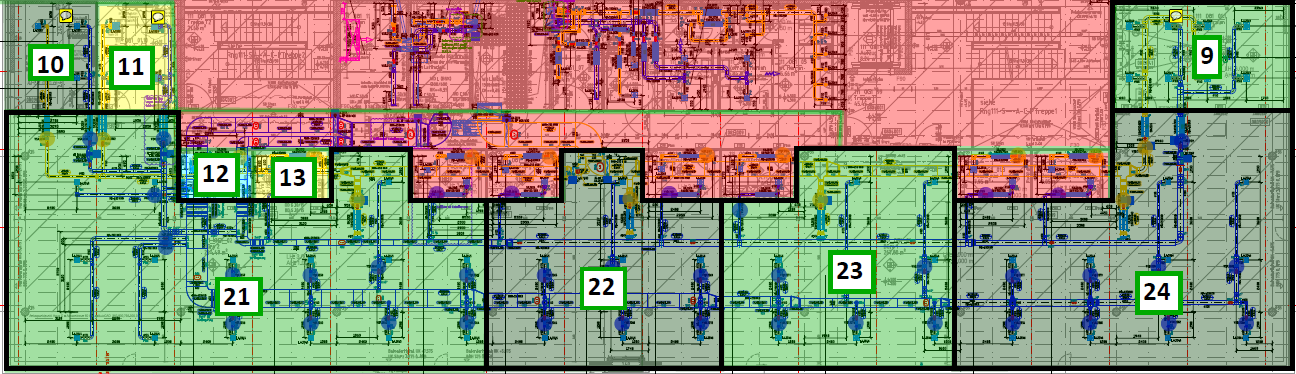}
	\caption{Zone layout of the considered part of the R\&D building at the Bosch Research Campus. Only zone 9 and zone 21 to 24 are considered.}
	\label{fig:layout}
\end{figure*}

This test case is based on an R\&D building at the Bosch Research Campus in Renningen. The zone layout is shown in Figure \ref{fig:layout}. The open space office is divided into four zones, namely zone 21-24, which are included in the simulation. Zone 9-11 are meeting rooms, where only zone 9 is considered due to low occupancy inside zone 10 and 11. Each zone is described by the model \eqref{eq:GB_Zone} with fitted parameters provided by Bosch.

The ground truth occupancy data is estimated with the UKF from CO\textsubscript{2} measurements, see again Figure \ref{fig:Occ}. Based on this ground truth, the hyperparameters of the GPs were optimized with the GPML Toolbox \cite{Rasmussen2010} over a horizon of 8 weeks. In the simulation, estimated occupancy data and measured ambient temperature from Bosch is used.

\subsection{MPC and Simulation Parameters}
The following simulation parameters are common between the case studies. The simulation is carried out for a whole year. The OCP \eqref{eq:OCP} with a prediction horizon $T = \qty{6}{\hour}$ is solved using the MPC toolbox GRAMPC \cite{Englert2019}. 
Only integral costs are used which are defined by
\begin{equation}
	l(T_z,\mb{u}) = \mb{u}^T\mb{R}\mb{u} + c\cdot h(T_z)
\end{equation}
with the weight $c = 0.01$, the control vector $\mb{u} = \left[\dot{m}_a, \: \dot{m}_w \right]^T $ and the penalty function
\begin{equation}
	h(T_z) = \begin{cases}
		0.5\cdot(T_z - T_u)^2 &\!\!\!\!\!\!, \;\; \text{if} \; T_z > T_u \\
		0.5\cdot(T_z - T_l)^2 &\!\!\!\!\!\!, \;\; \text{if} \; T_z < T_l \\
		0 &\!\!\!\!\!\!, \;\;  \text{otherwise}
	\end{cases}
	\label{eq:penalty}
\end{equation}%
\begin{table}
	\caption{Temperature comfort bounds.}
	\label{tab:bounds}
	\centering
	\begin{tabular}{l|l|l|l}
		Mode & Comfort & Pre-Comfort & Economy\\
		\hline
		$T_u$ & \qty{24}{\degreeCelsius} & \qty{25}{\degreeCelsius} & \qty{28}{\degreeCelsius}\\
		$T_l$ & \qty{21}{\degreeCelsius} & \qty{19}{\degreeCelsius} & \qty{17}{\degreeCelsius} \\
		\hline
	\end{tabular}
\end{table}%
for soft constraining the zone temperature with an upper and lower bound $T_u$ and $T_l$.

The MPC uses different modi to account for the occupancy, for which $T_l$ and $T_u$ in \eqref{eq:penalty} are chosen as shown in Table \ref{tab:bounds}. The Economy mode is used during weekends and at night from 6~pm to 6~am. The Comfort mode is statically used from 6~am to 6~pm or dynamically if occupancy is detected. The Pre-Comfort mode is used, if no occupancy is detected. 

These comfort modes are used in different controller scenarios, which are shown in Table \ref{tab:Modes}. The scenarios use different occupancy prediction schemes combined with different comfort mode selection. The \textbf{None} scenario represents the current state of the art, which the other controllers are compared to.
The \textbf{Exact} scenarios works as a baseline to show the best possible performance, as it has perfect knowledge about occupancy.
The threshold for switching between Pre-Comfort and Comfort is $0.5$ people, to account for rising slopes of the estimated occupancy data.

The controller scenarios are compared for thermal discomfort and energy consumption. An illustrative way to quantify thermal discomfort is the integration of the constraint violations, resulting in Kelvin hours $\left[ \text{Kh}\right]$. The energy consumption of the zone is calculated for the Renningen test case, whereby EnergyPlus provides the energy consumption of the whole building. 

\begin{table}[h]
	\centering
	\caption{Considered controller scenarios.}
	\label{tab:Modes}
	\begin{tabular}{p{2cm}|p{6cm}}
		\hline
		\textbf{None} & The MPC has no knowledge of the occupancy and switches between Economy and Comfort mode.\\ \hline
		\textbf{Exact} & Similar to \textbf{None}, but the MPC has exact knowledge about the occupancy.\\ \hline
		\textbf{LFM} & Similar to \textbf{None}, but the occupancy is predicted by the LFM.\\ \hline
		\textbf{Exact + Pre-Comfort} &  The MPC has exact knowledge of the occupancy. The default mode during the day is Pre-Comfort. If occupancy is detected, Comfort is used.\\ \hline
		\textbf{LFM + Pre-Comfort} & Similar to \textbf{Exact + Pre-Comfort}, but the occupancy is predicted by the LFM.\\
		\hline
	\end{tabular}
\end{table}

Both test cases were simulated in Python with an interface to GRAMPC\footnote{https://github.com/grampc/pygrampc} and own implementations of the Kalman Filter. Additionally, an UKF is used for state estimation in the EnergyPlus test case as only the room temperature is available.

\subsection{EnergyPlus Simulation Results}

\begin{figure}[h]
	\centering
	\definecolor{mycolor1}{rgb}{0.00000,0.44700,0.74100}%
\definecolor{mycolor2}{rgb}{0.85000,0.32500,0.09800}%
\newcommand{\barwidth}{20pt}
\begin{tikzpicture}
	
	\begin{axis}[%
		width=7.2cm,
		height=4cm,
		at={(0,0)},
		scale only axis,
		bar shift auto,
		symbolic x coords={Exact,LFM,Exact Adapt,LFM Adapt,None},
		enlarge x limits=0.2,
		xtick=data,
		xticklabel style={align=left},
		xticklabels = {,,,,},
		ymin=0,
		ymax=700,
		ylabel style={font=\footnotesize},
		ylabel={Discomfort in Kh},
		axis background/.style={fill=white},
		bar width=10pt,
		xmajorgrids,
		ymajorgrids,
		yminorgrids,
		minor tick num=3,
		minor grid style={gray!20},
		legend style={at={(0.0125,2.9cm)}, anchor=south west, legend cell align=left, align=left, draw=white!15!black}
		]
		\addplot[ybar, fill=mycolor1, draw=black, area legend] coordinates{%
			(Exact,	486.64054)
			(LFM,	484.17548)
			(Exact Adapt,	443.22126)
			(LFM Adapt,	437.95160)
			(None,	652.02832)
		};
		\addlegendentry{South}
		\addplot[ybar, fill=mycolor2, draw=black, area legend] coordinates{%
			(Exact,	345.47131)
			(LFM,	333.44347)
			(Exact Adapt,	332.32580)
			(LFM Adapt,	311.50504)
			(None,	236.36621)
		};
		\addlegendentry{North}
	\end{axis}
	
	\begin{axis}[%
		width=7.2cm,
		height=4cm,
		at={(0,-4.5cm)},
		scale only axis,
		bar shift auto,
		symbolic x coords={Exact,LFM,Exact Adapt,LFM Adapt,None},
		enlarge x limits=0.2,
		xtick=data,
		xticklabel style={font=\footnotesize,align=left},
		xticklabels = {Exact,LFM,Exact +\\ Pre-\\Comfort,LFM +\\ Pre-\\Comfort,None},
		ymin=0,
		ymax=15,
		ylabel style={font=\footnotesize},
		ylabel={Energy reduction in \%},
		axis background/.style={fill=white},
		xmajorgrids,
		ymajorgrids,
		yminorgrids,
		bar width=\barwidth,
		minor grid style={gray!20},
		minor y tick num=4
		]
		\addplot[ybar, fill=mycolor1, draw=black, area legend] coordinates{%
			(Exact,	4.062)
			(LFM,	4.009)
			(Exact Adapt,	13.789)
			(LFM Adapt,	12.732)
			(None, 0)
		};
	\end{axis}
	
\end{tikzpicture}%
	\caption{Energy reduction and thermal discomfort for the EnergyPlus test case. The thermal discomfort is broken down to the two offices in the north and south.}
	\label{fig:EPlusResults12h}
\end{figure}
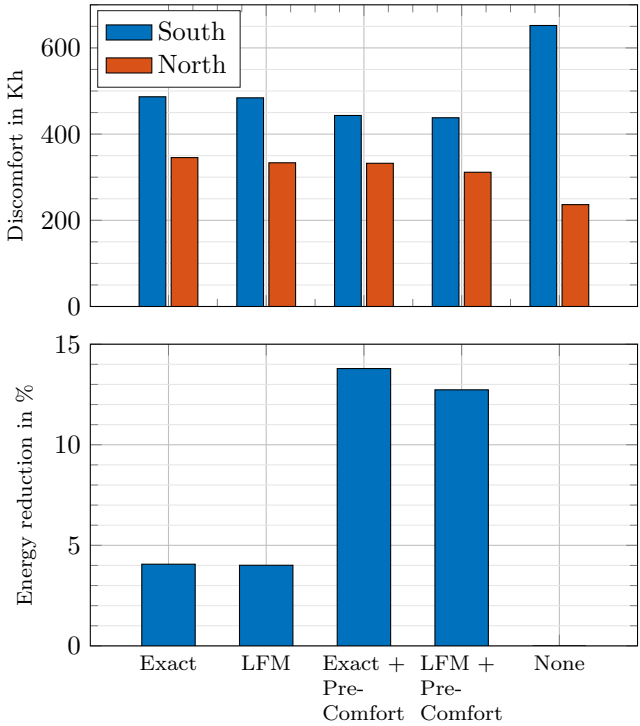

The results for the EnergyPlus test case are presented in Figure \ref{fig:EPlusResults12h}. The controller scenario \textbf{Exact} results in a lower discomfort for the south office, and a higher discomfort for the north office compared to the \textbf{None} scenario. This is due to disturbances like solar radiation and model errors. The south office is subject to solar radiation, which introduces a heat gain. This heat gain results in higher discomfort, especially during summer. In the winter, this heat gain helps to meet the comfort bounds. If the MPC additionally has information of the occupancy, which is another heat source, it can drive the zone closer to the comfort bounds. Additional disturbances can benefit, as can be seen in the south office, or harm this procedure, which can be seen in the north office. Additionally, the \textbf{Exact} scenario results in the highest discomfort for every other scenario excluding \textbf{None}. \textbf{LFM} achieves a similar performance as \textbf{Exact}, which shows that the LFM can accurately predict the occupancy. The Pre-Comfort scenarios result in a lower discomfort, with \textbf{LFM + Pre-Comfort} outperforming \textbf{Exact + Pre-Comfort}. This is due to the LFM smoothing the slopes of the occupancy data. Therefore, the Comfort mode is used sooner than in the \textbf{Exact + Pre-Comfort} scenario and thus results in a lower discomfort. 

The energy demand is reduced by $4\%$ for \textbf{Exact} and $4\%$ for \textbf{LFM}. 
This shows a comparable performance for \textbf{LFM}.
If the pre-comfort mode is used, a higher energy reduction can be achieved. For \textbf{Exact + Pre-Comfort} this results in $14\%$ and $12.5\%$ for \textbf{LFM + Pre-Comfort}. Here, the exact prediction achieves a higher energy reduction, because minimum time is spent in the Comfort mode. As said before, the LFM switches sooner to Comfort and therefore results in a lower reduction as more energy has to be spend to account for the higher temperature bounds.

This shows that a MPC with occupancy information can reduce the energy demand of a building. Even with additional disturbances like solar radiation this result holds. 
The reduction of discomfort depends directly on the constraint violation. If the MPC is tuned less conservatively, the constraints are more likely to be violated in the presence of model errors. In general, additional disturbances aggravate this effect. However, a positive impact of disturbances can be seen in comparison between the south and north office. In this case, the solar radiation mitigates the model error in the south office, resulting in lower constraint violation compared to the north office.

\subsection{Renningen Occupancy Prediction}
\label{sec:RenOcc}

Before discussing the MPC simulation results for the Renningen test case, the occupancy prediction quality is evaluated. As mentioned in Section 3, five individual GPs are used for each weekday. For every time step, the Root Mean Square Error (RMSE) is computed for the prediction and is summed up for the whole year. The results are shown in Table 3. One can see that the LFM achieves a lower error for every considered zone. For the zone 9, 23 and 24, the RMSE was not lowered as much as for zone 21 and 22. This is due to zone 9 being a meeting room with more irregular occupancy patterns and zone 23 and 24 with a considerably lower occupancy than zone 21 and 22.

\begin{table*}
	\centering
	\caption{Summed prediction RMSE of a whole year for the occupancy data from Renningen.}
	\label{tab:RMSE}
	\begin{tabular}{l|r|r|r|r|r}
		Zone & Zone 9 & Zone 21 & Zone 22 & Zone 23 & Zone 24 \\
		\hline
		LFM prediction & 12745 & 31900 & 17230 & 11364 & 12871 \\
		no prediction & 13830 & 54459 & 31119 & 12987 & 13884 
	\end{tabular}
\end{table*}

\subsection{Renningen Simulation Results}

\begin{figure}[t]
	\centering
	\definecolor{mycolor1}{rgb}{0.00000,0.44700,0.74100}%
\definecolor{mycolor2}{rgb}{0.85000,0.32500,0.09800}%
\definecolor{mycolor3}{rgb}{0.92900,0.69400,0.12500}%
\definecolor{mycolor4}{rgb}{0.494,0.184,0.55600}%
\definecolor{mycolor5}{rgb}{0.46600,0.67400,0.18800}%
\newcommand{\barwidth}{4.5pt}
\begin{tikzpicture}
	
	\begin{axis}[%
		width=7.2cm,
		height=4cm,
		at={(0,0)},
		scale only axis,
		bar shift auto,
		symbolic x coords={Zone 9,Zone 21,Zone 22,Zone 23,Zone 24},
		enlarge x limits=0.2,
		xtick=data,
		xticklabels = {,,,,},
		ymin=0,
		ymax=550,
		ytick={0, 100, 200, 300, 400, 500},
		ylabel style={font=\footnotesize\color{white!15!black}},
		ylabel={Discomfort in Kh},
		axis background/.style={fill=white},
		bar width=\barwidth,
		xmajorgrids,
		ymajorgrids,
		yminorgrids,
		minor tick num=1,
		minor grid style={gray!20},
		legend columns=5, 
		legend style={at={(-2.5cm,4.5cm)}, anchor=south west, legend cell align=left, align=left, draw=white!15!black}
		]
		\addplot[ybar, fill=mycolor1, draw=black, area legend] coordinates{%
			(Zone 9,	125.55012)
			(Zone 21,	447.40774)
			(Zone 22,	240.30719)
			(Zone 23,	148.20653)
			(Zone 24,	254.20954)
		};
		\addlegendentry{Exact}
		
		\addplot[ybar, fill=mycolor2, draw=black, area legend] coordinates{%
			(Zone 9,	246.84488)
			(Zone 21,	459.00578)
			(Zone 22,	257.69083)
			(Zone 23,	180.74127)
			(Zone 24,	265.00857)
		};
		\addlegendentry{LFM}
		
		\addplot[ybar, fill=mycolor3, draw=black, area legend] coordinates{%
			(Zone 9,	91.39723)
			(Zone 21,	404.88837)
			(Zone 22,	225.26491)
			(Zone 23,	111.97394)
			(Zone 24,	196.63573)
		};
		\addlegendentry{Exact \\ Pre- \\ Comfort}
		
		\addplot[ybar, fill=mycolor4, draw=black, area legend] coordinates{%
			(Zone 9,	319.93437)
			(Zone 21,	484.85769)
			(Zone 22,	368.65773)
			(Zone 23,	455.68030)
			(Zone 24,	513.25572)
		};
		\addlegendentry{LFM \\ Pre- \\ Comfort}
		
		\addplot[ybar, fill=mycolor5, draw=black, area legend] coordinates{%
			(Zone 9,	322.08529)
			(Zone 21,	453.27979)
			(Zone 22,	383.20553)
			(Zone 23,	177.04565)
			(Zone 24,	263.79302)
		};
		\addlegendentry{None}
	\end{axis}
	
	\begin{axis}[%
		width=7.2cm,
		height=4cm,
		at={(0,-4.5cm)},
		scale only axis,
		bar shift auto,
		symbolic x coords={Zone 9,Zone 21,Zone 22,Zone 23,Zone 24},
		enlarge x limits=0.2,
		xtick=data,
		ymin=0,
		ymax=15,
		ylabel style={font=\footnotesize},
		xticklabel style={font=\footnotesize},
		ylabel={Energy reduction in \%},
		axis background/.style={fill=white},
		xmajorgrids,
		ymajorgrids,
		yminorgrids,
		minor y tick num=4,
		minor grid style={gray!20},
		bar width=6pt,
		]
		\addplot[ybar, fill=mycolor1, draw=black, area legend] coordinates{%
			(Zone 9,	2.94213)
			(Zone 21,	2.60361)
			(Zone 22,	2.53772)
			(Zone 23,	0.84134)
			(Zone 24,	0.39901)
		};
		
		\addplot[ybar, fill=mycolor2, draw=black, area legend] coordinates{%
			(Zone 9,	0.51023)
			(Zone 21,	2.11157)
			(Zone 22,	2.21942)
			(Zone 23,	0.43012)
			(Zone 24,	0.16794)
		};
		
		\addplot[ybar, fill=mycolor3, draw=black, area legend] coordinates{%
			(Zone 9,	13.43439)
			(Zone 21,	6.27991)
			(Zone 22,	5.58651)
			(Zone 23,	7.77916)
			(Zone 24,	8.98976)
		};
		
		\addplot[ybar, fill=mycolor4, draw=black, area legend] coordinates{%
			(Zone 9,	10.05244)
			(Zone 21,	5.82240)
			(Zone 22,	5.29295)
			(Zone 23,	8.97405)
			(Zone 24,	9.58830)
		};
	\end{axis}
	
\end{tikzpicture}%
	\caption{Energy reduction and thermal discomfort for the Renningen test case.}
	\label{fig:RenningenResults6h}
\end{figure}
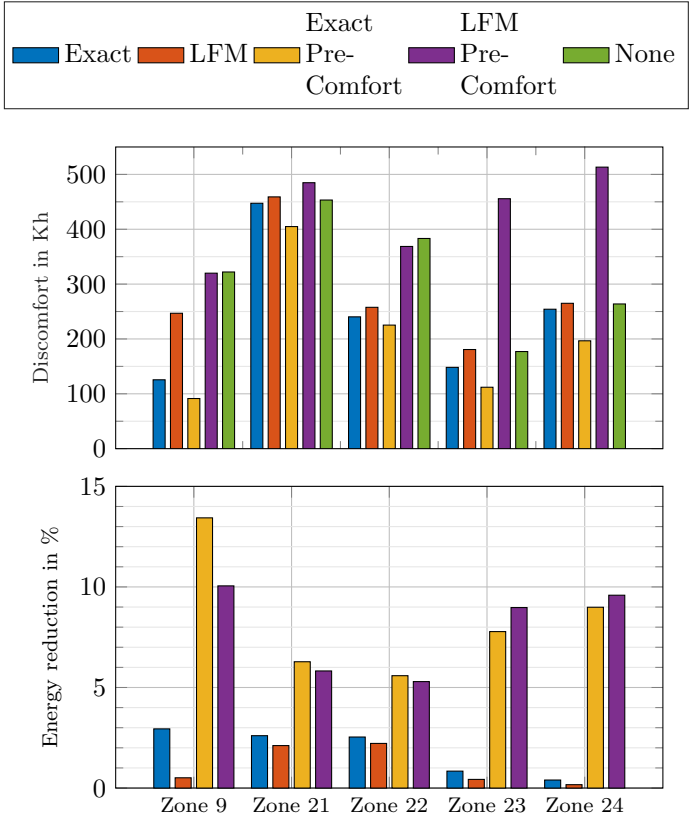

The results for the Renningen test case differ from the EnergyPlus case, because occupancy data based on real measurements are used. They are presented in Figure \ref{fig:RenningenResults6h}. There is much more variation in the thermal discomfort. The \textbf{Exact} and \textbf{Exact + Pre-Comfort} scenarios achieve the lowest discomfort as expected. Differently than in the EnergyPlus test case, the simulation model was \eqref{eq:GB_Zone} with occupancy information. Therefore, the MPC used an exact model with no further unknown disturbances. Hence the discomfort is lower.

The \textbf{LFM} scenario shows a moderate higher discomfort than \textbf{Exact} for zone 21 to 24. For zone 9 the discomfort is considerably higher than \textbf{Exact}, but still lower than \textbf{None}. This shows that the prediction quality of the LFM is good enough to achieve a lower discomfort, which is consistent with the results in Section \ref{sec:RenOcc}.

For zone 9 and 22 the \textbf{LFM + Pre-Comfort} scenario works as expected, as it reduces the thermal discomfort. For zone 21 a marginal higher discomfort than \textbf{None} is achieved. But in zone 23 and 24, it leads to a twice as large discomfort as \textbf{None}. This is due to the LFM incorrectly predicting the occupancy and the MPC therefore predicting the wrong comfort mode. Additionally, if two consecutive weekdays vary greatly, the prediction quality of the LFM becomes significantly worse. \textbf{Exact + Pre-Comfort} doesn't has these problems, as the predictions are exact. Additionally, \textbf{Exact + Pre-Comfort} achieves a lower discomfort than \textbf{Exact}. This can be due to switching from Economy to Pre-Comfort at the start and end of the day, because the step of the switching constraints is lower. 

For zone 9, 21 and 22, \textbf{Exact} leads to an energy reduction by approx. $3\%$. The energy reduction for zone 23 and 24 is only $1\%$. This can be due to low occupancy, as the MPC can't exploit the additional heat induced by occupancy. \textbf{LFM} shows similar results for zone 21 to 24 like \textbf{Exact}. For zone 9 the energy reduction is much lower, as the LFM struggles to predict the occupancy correctly. The occupancy in zone 9 is especially hard to predict for the LFM. This is due to irregular scheduled meetings. Furthermore, meetings are often scheduled biweekly. 

In contrast, the \textbf{Pre-Comfort} scenarios achieve considerably higher energy reductions, especially for zone 23 and 24. Furthermore, \textbf{LFM + Pre-Comfort} results in a higher energy reduction at the cost of higher discomfort. For the remaining zones, \textbf{Exact + Pre-Comfort} achieves a higher energy reduction. Only for zone 9, an energy reduction of about $13\%$ is achieved. 

If the MPC only exploits the heat gain of occupants, the energy reduction is lower than adapting the comfort bounds. Additionally, one can see that \textbf{Exact + Pre-Comfort} and \textbf{LFM + Pre-Comfort} have the same difference as \textbf{Exact} and \textbf{LFM}. Therefore it can be assumed that the potential for exploiting the heat gain in this case study is about $3\%$, if the occupancy is high enough.

	\section{Conclusion} 
\label{sec:conclusion}

This paper shows a new method to predict occupancy. A GP is used to learn the occupancy pattern, which is then represented as a state space model. This enables the use of Kalman filtering for prediction and incorporating new data. Furthermore, if the state dimension is lower than the number of data points used in the GP, Kalman filtering is more efficient. The prediction trajectory for a single time step is pre-computed and used inside an MPC for BES. 

Generally incorporating occupancy data for MPC has a positive impact on reduced energy demand. The energy reduction ranges from 3\% to 14\%, depending on the considered zone. The additional occupancy information results in a less conservative MPC. However, unconsidered disturbances or model errors can lead to a higher or lower discomfort. If only the heat gain of occupants is included, the LFM achieves a similar performance as with exact occupancy information. 

Including an additional comfort mode leads to further energy reductions. The Pre-Comfort mode is used if no occupancy is detected. For dense occupied zones these reductions are at least twice as large than using only the heat gain of the occupants. The potential is greater for primary unoccupied zones like meeting rooms. 

Overall, this paper shows the applicability of the LFM approach for occupancy prediction. Especially the results of the EnergyPlus test case reveals the possible performance with model errors and additional disturbances.

Potential future work includes the kernel selection. In this paper, the choice of the kernel function was determined a~priori through observation of the data. This task is complex because prior knowledge of the data features must be available to choose an appropriate kernel. Choosing a general kernel which can approximate arbitrary kernel functions can simplify this task. Furthermore, online adaption of the hyperparameters can be considered to react on changing features of the data.

	\section{Acknowledgements}
	This work was supported by the German Federal Ministry for Economic Affairs and Climate Action (BMWK) with promotional reference 03EN1066B.
	
	\appendix
	\section{Supplementory Data}
	The estimated occupancy data from Renningen with an implementation of the LFM prediction method is available at https://github.com/ThoreWietzke/occupancy-benchmark-dataset
		
	\bibliographystyle{elsarticle-num} 
	\bibliography{ref_no_url}

\end{document}